\documentclass[twocolumn,amssymb,amsmath,amstext,aps]{revtex4}
\usepackage{graphicx}
\usepackage{dcolumn}
\usepackage{bm}
\usepackage{latexsym,epsf,epsfig}
\usepackage{array}
\usepackage{caption}
\usepackage{amstext}
\usepackage[pdfborder=0 0 0]{hyperref}

\begin{document}

\title{Brief remarks on ``Electric dipole moment enhancement factor of thallium''}
\author{H. S. Nataraj$^{1}$}\email{nataraj@cycric.tohoku.ac.jp}\author{B. K. Sahoo$^2$} \author{B. P. Das$^3$ } \author{D. Mukherjee$^4$}
\affiliation{$^1$ Cyclotron and Radioisotope Center, Tohoku University, 9808578 Sendai, Japan}
\affiliation{$^2$ Theoretical Physics Division, Physical Research Laboratory, Ahmedabad 380009, India}
\affiliation{$^3$ Theoretical Astrophysics Group, Indian Institute of Astrophysics, Bangalore 560034, India}
\affiliation{$^4$ Raman Center for AMO Sciences, Indian Association for the Cultivation of Science, Kolkata 700032, India}

\begin{abstract}
In a recent paper by Porsev {\em et al.} [arXiv:1201.5615v1], the authors have 
claimed to have resolved the controversy arising from the different {\it ab initio} 
results available for the EDM enhancement factor of Tl. In our opinion, any such 
attempt to resolve the discrepancies between different calculations has to 
thoroughly compare the basis sets used, methods employed and approximations 
considered in the two different cases. However, Porsev {\em et al.} have not 
succeeded in doing so in their current paper. We clarify some of their 
misunderstandings about our work and address some specific issues in this note.
\end{abstract}

\maketitle

First of all, we would like to state that some of the points made by Porsev 
{\em et al.} in their paper \cite{Porsev} are factually incorrect. We quote, for 
example, their statement: ``Since the valence-valence correlations are very large, 
the CI method provides better description of these correlations than the 
perturbative approaches such as RCC due to possible large contributions of 
higher-order (or higher-excitation) correlations.''. The fact is that the 
coupled-cluster (CC) theory, or its relativistic extension, RCC, is NOT perturbative
\cite{bishop,bartlett} even though it is equivalent to all order perturbation 
theory. Furthermore, the configuration interaction (CI) approach is a subset of 
CC method at the same level of hole-particle excitations and hence, it does not 
include any higher-order correlations beyond CC theory as they have claimed. For 
example, CC singles and doubles (CCSD) contains all the terms that are present in 
CI singles and doubles (CISD) and much more \cite{bishop,bartlett}. Further, we would like 
to clarify that the valence-valence correlations are very large in thallium (Tl), 
as they have reported, only if one treats $6s^2\,6p$ as valence electrons. However,
 that does not preclude one from considering $6s^2$ as part of the core and 
$6p_{1/2}$ as valence and treating the valence-core interactions to all order in the
residual Coulomb interaction via the CC theory as we have done \cite{nataraj1}.
Therefore the valence-valence correlation of Porsev {\em et al.} \cite{Porsev}
and Dzuba and Flambaum (D \& F) \cite{dzuba1} is a part of our valence-core 
correlation.

  Furthermore, the same many-body approach is employed by Porsev {\em et al.} 
\cite{Porsev} for their calculations using $V^{N-3}$ as well as $V^{N-1}$ 
orbitals. In both cases, they have treated the three outermost electrons as part 
of the valence space using CI and the rest of the electrons as core using MBPT.  
In this context, they mention that the RCC approach employed by us does not treat 
$6s\,6p^2$ and $6s^2\,ns$, where $n$ being $\ge 7$, on an equal footing. However, 
this is not true. We have treated both of them on the same footing, that is, as 
opposite parity excitations from a common reference state \cite{nataraj1}.

  Porsev {\em et al.} \cite{Porsev} are correct in stating that: ``It is very 
important to accurately account for the contributions of the $6s\,6p^2$ 
configurations.''. However due to their incorrect assumptions that we have 
mentioned above, they have arrived at an incorrect conclusion that ``In the RCC 
method (in referring to our work in \cite{nataraj1}), these contributions were treated as excitations
of the core electrons which is unlikely to provide the required accuracy.''  They 
imply that $6s^2$ electrons has to be treated only as valence, or in other words, 
one has to treat Tl only as a three-valence atom as against our treatment of it 
as a monovalent atom. Referring to Tl as a one-valence or three-valence system is 
a matter of semantics. However, what really matters is not the terminology one 
uses (three-valence approximation vs. one-valence approximation), but rather the 
physical effects contained in a particular theory. Indeed, Porsev {\em et al.} 
\cite{Porsev} and D \& F \cite{dzuba1} have considered $6s^2$ and $6p_{1/2}$ as 
valence electrons and have evaluated the valence-valence correlations by CI. The 
net level of excitation of the configurations in their CI calculations does not
appear to be beyond triples with reference to $6s^2\,6p_{1/2}$. In contrast, our
CC theory has all linear and nonlinear single and double excitations from all
the core and valence electrons (in particular $6s^2$ and $6p_{1/2}$ electron)
and their products as given in Eqs. (3 \& 4) of our paper \cite{nataraj1}. In other 
words, in addition to single and double excitations from the $6s^26p_{1/2}$ 
Fermi vacuum state, our wave function consists of a large number of triple,  
quadruple, quintuple, sextuple, septuple and octuple excitations which are obtained
as disconnected products of single and double excitations. Therefore
in addition to all the configurations included in the calculations of Porsev
{\em et al.} \cite{Porsev} and D \& F \cite{dzuba1}, our
calculation contains many more configurations corresponding mostly to higher 
order excitations. From the information given on the configurations included
in the model space and the excitations considered for the CI calculation
of D \& F, it does appear that this calculation is not size extensive \cite{bishop,bartlett} 
as the CI is not complete in the chosen orbital space. Even though Porsev {\em et al.} do not specify 
their configurations in the model space, it seems as though they have not performed
a full CI, particularly in view of their statement ``higher n orbitals were 
allowed fewer number of excitations'', and therefore it also might not be size 
extensive. In contrast, our RCC calculation is size extensive at all levels of 
excitation. The probable lack of size extensivity in the two above mentioned CI 
calculations could be one of the likely sources of discrepancy between those
calculations and ours.

   Further, we also would like to clarify that contrary to the opinion expressed 
by Porsev {\em et al.}, the size of our basis set is not small. If our basis 
functions were not nearly complete then we would not have obtained good results 
for both the electric dipole (E1) amplitude and the magnetic dipole hyperfine 
constants shown together in Table IV of our paper \cite{nataraj1}. Very 
surprisingly, Porsev {\em et al.} make no reference at all to these results. The 
wave functions at both near- and far-nuclear regions are subjected to scrutiny 
through these properties. On the whole our results are in better agreement with 
experiments than those of D \& F's and Porsev {\em et al.}'s results; For 
example, our RCC method gives E1 amplitude of the $7S \rightarrow 6P_{1/2}$ 
transition as 1.82 au against the experimental value 1.81(2) au. In contrast,
this is reported as 1.781 au and 1.73 au by Porsev {\em et al.} \cite{Porsev}
and D \& F \cite{dzuba1}, respectively, which are outside of experimental limit.
The good agreement with experiments of our results for the above mentioned two 
properties which are related to the EDM enhancement factor calculation is a 
strong indication that our EDM enhancement factor is reliable. In this context,
we would also like to mention that we have considered $38s; 34p_{1/2;3/2};
34d_{3/2;5/2}; 30f_{5/2;7/2}$ and $20g_{7/2;9/2}$ number of orbitals for the
SCF calculation. The number of basis functions and the basis parameters in the
finite basis space are well optimised by comparing the single particle energies 
and SCF energy with respect to the numerical (non-parametrical) results obtained 
by the general-purpose relativistic atomic structure program (GRASP). We have
performed various tests at the Dirac-Fock level to check the correctness of our single 
particle orbitals like the comparison of single particle energy differences, 
dipole and EDM matrix elements with those obtained using GRASP (details can be
found in \cite{natthes}). After the SCF step, we have truncated the virtual space 
by dropping the high-lying virtuals whose contribution to the results is small. 
As a matter of fact, those basis functions which are used by Porsev {\em et al.}, 
B-splines used by D \& F and Gaussians used in our work have different qualitative
behaviours at different regions. Therefore, one cannot directly compare the number 
of basis functions used in different calculations. Having said that, we would 
like to emphasize that the use of Gaussian basis functions in relativistic 
calculations have been well tested for a number of different properties.

  We also emphasize that the EDM enhancement factor for Tl depends mainly on the 
following three factors: the EDM matrix elements, the E1 matrix elements and the 
energy differences between the ground state and the intermediate states. 
Porsev {\em et al.} have calculated the $6p_{1/2} - 7s$ EDM matrix element using 
the CC method restricting the number of basis to $n = 14$ for all partial waves 
in an attempt to reproduce our result and have reported that it reduces the value 
by 18\% when compared to their calculation with a bigger basis set considered for 
the rest of the calculations. However, this reduction will not be reflected in the 
total EDM enhancement factor by a similar magnitude. As the energy differences 
between the high-lying virtual states and the ground state appearing in the 
denominator will be large, the contribution from those states to the overall EDM enhancement factor would be small.

  Porsev {\em et al.} have calculated the core-valence correlations using the 
CI$+$all-order method only in the $V^{N-3}$ case and they report that their 
contribution to the total EDM enhancement factor is less than 1\%. Further, they 
assume that the magnitude of these correlations will more or less be the same in the $V^{N-1}$ 
case. As the nature of the orbitals are very different in the two cases, for details please see \cite{nataraj2}, this assumption may not be valid. In Ref. \cite{nataraj2} we have demonstrated by numerical calculations that the EDM enhancement factor of Tl computed using the $V^{N-3}$ orbitals over-estimates the result in comparison to that calculated using the $V^{N-1}$ orbitals at the Dirac-Fock level of the theory.
Further, their CI$+$all-order approach only includes the linearized CC terms and 
hence, the contributions from the non-linear CC terms which are omitted in their 
calculations may also have non-negligible contributions.

   Porsev {\em et al.} have used the sum-over-states approach to calculate the 
EDM enhancement factor from the two specific states: $6s^2\,7s$ and $6s^2\,8s$ 
using the RCC method and they compare those results with their full 
CI$+$MBPT$+$RPA results and both these results agree to each other, within 2\% 
accuracy. Further they have commented that the result inferred from the Figure 2 
of our paper is 10\% lower than their RCC result for the $6s^2\,7s$ state. However, we would like to 
remind the fact that we had given the combined contributions of the following RCC 
terms: $DT_1^{(1)}$, $DS_{1v}^{(1)}$, $S_{1v}^{(0)\dagger} D S_{1v}^{(1)}$ and 
$S_{2v}^{(0)\dagger} D S_{1v}^{(1)}$ to the singly-excited intermediate states of 
$s$ symmetry in Figure 2 of our paper \cite{nataraj1}, as quoted in the text 
therein. It is not clear to us whether or not Porsev {\em et al.} have included 
the contributions from all these RCC terms in their results and if not, then the 
comparison may not be very meaningful. 

 On a note aside, we remark that the calculation of the effective electric 
field of a molecule such as YbF is different from the calculation of the EDM 
enhancement factor of Tl on the following grounds: the question of one- or 
three-valence does not arise in the former case and in addition, the amount of 
correlation to the effective electric field of YbF is very small, about 3\%, as 
observed in all the earlier calculations including our own recent RCC calculation,
unlike the case of Tl where the correlation contributions are large. So, we do 
not agree with the view expressed by Porsev {\em et al.} that the YbF calculations
are more difficult than the Tl calculations.


 In summary, the current attempt by Porsev {\em et al.} \cite{Porsev} in trying to 
resolve the discrepancies between different EDM enhancement factor values for Tl has merely
added a new result to the literature. Although, it reports an agreement, within 
the limits of the quoted uncertainties, with the earlier two calculations 
\cite{dzuba1,liu}, it clearly emphasizes that the correlation effects ignored by 
D \& F in their CI+MBPT+RPA calculation could be very important and it further quantifies the magnitude of these corrections to be as large as 7\% in the $V^{N-3}$ potential itself, despite the severe accidental cancellations among the different corrections. Porsev {\em et al.} also observe that the energies and various properties computed using the orbitals generated in the $V^{N-1}$ potential agree quite well with the experiments when compared to those results of $V^{N-3}$ potential in concurrence with our observation reported in \cite{nataraj2}. Fortuitously the net result of Porsev {\em et al.} agrees with the results of \cite{dzuba1,liu}. We observe that the method followed in the work of Porsev 
{\em et al.} \cite{Porsev} is not very different from that of  D \& F \cite{dzuba1} and 
their RCC calculation of the EDM enhancement factor is not as complete as ours. 
In the absence of adequate information about various calculations performed by 
Porsev {\em et al.}, particularly their CI calculation and a detailed comparison
between their intermediate results and ours given in Table I of our paper \cite{nataraj1},
it is difficult at this stage for us to judge their work.
However, due to the reasons discussed in this note we believe that Porsev 
{\em et al.}'s work on the EDM enhancement factor calculation of Tl as reported 
in \cite{Porsev} has come nowhere close to resolving the discrepancies between 
the different calculations.

\end{document}